\documentclass{article}
\usepackage{amsmath}
\usepackage{amssymb}
\usepackage{amsfonts}

\input{tcilatex}

\begin{document}

\title{Pseudoclassical Mechanics for the spin 0 and 1 particles.}
\author{R. Casana$^{(1)}$\thanks{%
casana@ift.unesp.br}~, M Pazetti$^{(1)}$\thanks{%
mpazetti@ift.unesp.br}~, B.M. Pimentel$^{(1)}$\thanks{%
pimentel@ift.unesp.br}~, and J.S. Valverde$^{(1,2)}$\thanks{%
valverde@ift.unesp.br,valverde@stout.ufla.br} \\
$^{\mathit{\small (1)}}$\textit{{\small Instituto de F\'{\i}sica Te\'{o}%
rica, Universidade Estadual Paulista}} \\
\textit{{\small Rua Pamplona 145, CEP 01405-900, S\~{a}o Paulo, SP, Brazil}}%
\\
$^{\mathit{\small (2)}}$\textit{{\small Departamento de Ci\^{e}ncias Exatas,
Universidade Federal de Lavras}}\\
\textit{\small Caixa Postal 3037, CEP 37200-000, Lavras, MG, Brazil}}
\date{}
\maketitle

\begin{abstract}
We give an action for the massless spinning particle in pseudoclassical
mechanics by using grassmann variables. The constructed action is invariant
under $\tau $-reparametrizations, local SUSY and $O(N)$ transformations.
After quantization, for the special case $N=2$, we get an action which
describes the spin 0, 1 particles and topological sectors of the massless
DKP theory. A SUSY formulation of the given model is also explored.
\end{abstract}

\section{Introduction}

Investigations of particle systems with arbitrary spin was initially given
by Bargmann-Wigner \cite{Bargmann} and Rarita-Schwinger \cite{Rarita}, here
the Dirac representations of the spin one half particles are the basis to
the construction of higher spin theories. The formalism is based on the
bispinor wave function with $2s$ Dirac indices (for spin $s$) and the total
symmetrical representation is used to study the maximum spin value of the
model.

On the other hand, the first ideas about the studies of classical
systems that include in the phase space both commuting and
anticommuting variables (pseudoclassical mechanics) was put forward
by Schwinger \cite{Schwinger} in 1953. However it was Martin
\cite{Martin} who achieved these ideas in 1959. Later in the Berezin
and Marinov works \cite{Berezin} a model for the description of spin
one half particles was proposed, here the consistent formulation of
the relativistic particle dynamics implies in the addition of a new
constraint, this is because the formulation of the massive case has
five grassmann variables. At the same time these models were also
studied by Casalbuoni \cite{Casalbuoni} who explored the internal
group symmetry and the gauge invariance of the resulting action. In
this way was possible the description of spinless and spin one
particles using these internal symmetries. Interaction of spinning
particle systems with external Yang-Mills and gravitational fields
was investigated in \cite{Casalbuoni1}. The quantization of similar
models are performed by means of the Dirac procedure for constrained
systems.

Many other papers appeared about the study of spinning particles in the
framework of pseudoclassical mechanics, for example the derivation of the
equation of motions for the massive and massless spinning particles are
treated in the works \cite{Brink1, Brink, Gershun1, Gershun2}, where the
spin description is achieved by means of the inclusion of internal group
symmetries. Similarly, the case of the Dirac particle is discussed in the
works \cite{Galvao, Gitman, Zlatev}. A path integral representation for
obtaining a Dirac propagator was also obtained in \cite{Fainberg1} and other
studies connecting the pseudoclassical mechanics with the string theory was
investigated \cite{Marshakov} for the free case as {in interacting with an
external field}. Also, the pseudoclassical description of massless Weyl
fermions and its path integral quantization when coupled to Yang-Mills and
gravitational fields was studied in \cite{new1}. Similarly, the path
integral quantization of spinning particles interacting with external
electromagnetic field was analyzed in \cite{new2}.

Besides this, the pseudoclassical approach can be applied to other different
models. This is the case of the Duffin-Kemmer-Petiau (DKP) theory \cite%
{Duffin, Kemmer, Petiau} which describes massive spin $0$ and spin $1$
particles in a unified representation. Questions about the equivalence of
the DKP theory with theories like Klein-Gordon and Maxwell are discussed in
\cite{Fainberg0, Fainberg01, Fainberg2} (a good historical review of the DKP
theory can be found in \cite{Krajcik, Krajcik1}). {The Field theory of the
massless DKP has a local gauge symmetry which describes the electromagnetic
field in its spin 1 sector. It is important to notice that the massless case
can not be obtained through the limit} $m\rightarrow 0$ {of the massive
case. This is due to the fact that the projections of DKP field into spin 0
and 1 sectors involve the mass as a multiplicative factor \cite{Casana1} so
that taking the limit} $m\rightarrow 0$ {makes the results previously
obtained useless. Moreover, if we simply make mass equal to zero in the
usual massive DKP Lagrangian we obtain a Lagrangian with no local gauge
symmetry}. Studies in the Riemann-Cartan space time was proposed in \cite%
{Casana, Casana0,Casana1}.

Recently, a super generalization of the DKP algebra was done by Okubo \cite%
{Okubo} where the starting point is the study of all irreducible
representations by means of the Lie algebra $so(1,4)$ \cite{Fishbach},
moreover, a paraDKP (PDKP) algebra is constructed intimately related to the
Lie superalgebra $osp(1,4),$ obtaining as result the super DKP algebra that
contains the boson and fermion representations.

An extended variant including Grassmann variables for the DKP theory is very
interesting for many reasons, for example a pseudoclassical version allow us
to make an attempt to the construction of a supersymmetry variant of the
theory where the action must be expressed in terms of (super)fields. It is
also no clear about the particle states that will compose the
(super)multiplet in this theory.

In this work we propose a possible action for the massless DKP theory in the
pseudoclassical approach. In section \textbf{2}, the pseudoclassical action
is given including the correct boundary terms that yields a consistent
equations of motions. We carry out the constraint analysis of the system and
verify his invariance under $\tau $-reparametrizations, internal group $O(N)$
and SUSY transformations. We find the generators of corresponding
transformations and give the Pauli-Lubanski vector. In section \textbf{3},
the quantization is performed and proved that for the special case $N=2$ the
both sectors of spin 0 and spin 1 of the DKP theory appear. We get the
scalar and vectorial field as a first result, we also obtain the topological
field solutions correspondent to the both spin sectors. In Section \textbf{4}%
, using the SUSY principles we extend the proposed action to the Superspace
formalism obtaining a consistent result as in the pseudoclassical model.
Finally in section \textbf{5}, we give our conclusions and comments.

\section{Pseudoclassical Mechanics}

We start with the action in the first order formalism that considers an
internal group symmetry
\begin{equation}
S=\int\limits_{\tau _{1}}^{\tau _{2}}d\tau \left[ \left( \overset{.}{x}%
-i\chi \psi \right) p+\frac{e}{2}p^{2}+\frac{i}{2}\psi \dot{\psi }+\frac{i}{2%
}f\psi \psi \right] +\frac{i}{2}\psi \left( \tau _{2}\right) \psi \left(
\tau _{1}\right)  \label{p1}
\end{equation}
here $x_{\mu }$ is the space time coordinate, $p_{\mu }$ the auxiliary
momentum vector; $\psi _{\mu }^{k}\left( \tau \right) -k,l,...=1,2,...N$ are
the fermion coordinates, superpartner of $x_{\mu }\left( \tau \right) $, $%
\left( x_{\mu },\psi _{\mu }^{k}\right) $ is the multiplet of matter; $%
e\left( \tau \right) $ is the \textit{einbein}, his superpartner $\chi
_{k}\left( \tau \right) $ is the unidimensional gravitino; $f_{ik}\left(
\tau \right) =-f_{ki}\left( \tau \right) $ is the gauge field for internal
symmetry, $\left( e,\chi _{k},f_{ik}\right) $ is the supergravitational
multiplet on the world line.

The action (\ref{p1}) includes the correct boundary terms that guarantee the
consistence of the equations of motions for the grassmann variables. This is
because in the variational principle the fermionic canonical coordinates
have only one condition
\begin{equation}
\delta \left( \psi \left( \tau _{2}\right) +\psi \left( \tau _{1}\right)
\right) =0  \label{p2}
\end{equation}
for the other coordinates only the space time coordinate is restricted to
the condition
\begin{equation}
\delta x\left( \tau _{2}\right) =\delta x\left( \tau _{1}\right) =0
\label{p3}
\end{equation}
internal group indices in the case $N=2$ when $i,k=1,2$ are contracted by
means of symbol Kroeneker $\delta _{ik}$ (for the group $O\left( 2\right) $
and spin $1$) or Levi-Civita symbol $\epsilon _{ik}$ (for the group $%
Sp\left( 1\right) $ and spin $0$).

The lagrangian that follows from (\ref{p1}) is
\begin{equation}
\mathcal{L}=\left( \dot{x}-i\chi \psi \right) p+\frac{e}{2}p^{2}+\frac{i}{2}%
\psi \dot{\psi }+\frac{i}{2}f\psi \psi  \label{p4}
\end{equation}

It is possible to write the action (\ref{p1}) in a different way, for this
we perform the variation of $S$\ with respect to $p$, then we get the
following equation
\begin{equation}
p=-e^{-1}\left( \dot{x}-i\chi \psi \right)  \label{p4a}
\end{equation}
inserting this solution into (\ref{p1}) we obtain the second order formalism
of the action
\begin{eqnarray}
S &=&\int\limits_{\tau _{1}}^{\tau _{2}}d\tau \left[ -\frac{e^{-1}}{2}\left(
\dot{x}^{2}-2i\dot{x}\chi \psi -\left( \chi \psi \right) ^{2}\right) +\frac{i%
}{2}\psi \dot{\psi }+\frac{i}{2}f\psi \psi \right]  \notag \\
&&+\frac{i}{2}\psi \left( \tau _{2}\right) \psi \left( \tau _{1}\right)
\label{p4b}
\end{eqnarray}
then the lagrangian that follows from (\ref{p4b}) is
\begin{equation}
\mathcal{L=}-\frac{e^{-1}}{2}\left( \dot{x}^{2}-2i\dot{x}\chi \psi -\left(
\chi \psi \right) ^{2}\right) +\frac{i}{2}\psi \dot{\psi }+\frac{i}{2}f\psi
\psi  \label{p4c}
\end{equation}
the term $\left( \chi \psi \right) ^{2}=\chi _{i}\psi _{i}\chi _{k}\psi _{k}$
appears because an internal group symmetry $O(N)$ was introduced in the
theory.

Both formulations (\ref{p1}) and (\ref{p4b}) are equivalent and as we will
see later the constraint analysis gives the same result.

Equations of motions that follow from the action (\ref{p1}) result in
\begin{equation*}
p_{\mu }\psi _{k}^{\mu }=0,\quad \psi _{\mu i}\psi _{k}^{\mu }=0,\quad \dot{%
\psi }_{k}^{\mu }=-p^{\mu }\chi _{k}+f_{ik}\psi _{i}^{\mu },\quad \dot{p}=0
\end{equation*}
we can see that for a special case $e=1,$ $\chi =f=0$ we obtain the
solutions
\begin{equation*}
x_{\mu }\left( \tau \right) =x_{\mu }\left( 0\right) +p_{\mu }\tau ,\quad
\psi _{k}^{\mu }=const.
\end{equation*}

\subsection{Constraint Analysis}

Now we proceed to the constraint analysis of the theory. Using the
definition for the canonical momentum: $p_{a}=\frac{\overleftarrow{\partial }%
\mathcal{L}}{\partial \dot{q}^{a}},$ we obtain
\begin{eqnarray}
p_{\mu } &=&\frac{\partial \mathcal{L}}{\partial \dot{x}^{\mu }}=p_{\mu
};\quad \pi _{\mu }^{k}=\frac{\partial \mathcal{L}}{\partial \overset{.}{%
\psi }_{k}^{\mu }}=\frac{i}{2}\psi _{\mu }^{k}  \label{p5} \\
\pi &=&\frac{\partial \mathcal{L}}{\partial \dot{e}^{\mu }}=0;\quad \pi ^{k}=%
\frac{\partial \mathcal{L}}{\partial \dot{\chi }_{k}}=0;\quad \pi ^{ik}=%
\frac{\partial \mathcal{L}}{\partial \dot{f}_{ik}}=0  \notag
\end{eqnarray}
from which a set of primary constraints appears
\begin{eqnarray}
\Omega _{\mu }^{k} &=&\pi _{\mu }^{k}-\frac{i}{2}\psi _{\mu }^{k}\approx
0,\quad \Omega _{\pi }=\pi \approx 0,\quad \Omega ^{k}=\pi ^{k}\approx
0,\quad \Omega ^{ik}=\pi ^{ik}\approx 0  \notag \\
&&  \label{p6}
\end{eqnarray}
following the standard Dirac procedure for a theory with constraints we
construct the primary hamiltonian from the lagrangian (\ref{p4}), $\mathcal{H%
}=p_{a}\dot{q}^{a}-\mathcal{L}$,
\begin{equation}
\mathcal{H}^{(1)}=i\chi _{k}\psi _{k}^{\mu }p_{\mu }-\frac{e}{2}p^{2}-\frac{i%
}{2}f_{ik}\psi _{\mu i}\psi _{k}^{\mu }+\lambda ^{a}\Omega _{a}  \label{p6a}
\end{equation}
where we have included the primary constraints (\ref{p6}), $\lambda
^{a}=\left\{ \lambda _{\mu }^{k},\lambda _{\pi },\lambda ^{k},\lambda
^{ik}\right\} $ are the lagrange multipliers. When we apply the stability
conditions on the primary constraints \
\begin{equation}
\dot{\Omega }_{a}=\left\{ \Omega _{a},\mathcal{H}^{(1)}\right\} _{PB}=0
\label{p7}
\end{equation}
we obtain a new set of secondary constraints
\begin{equation}
\Omega _{\pi }^{(2)}=\frac{1}{2}p^{2}\approx 0,\quad \Omega _{k}^{(2)}=i\psi
_{k}^{\mu }p_{\mu }\approx 0,\quad \Omega _{ik}^{(2)}=i\psi _{\mu i}\psi
_{k}^{\mu }\approx 0  \label{p8}
\end{equation}
the conservation of these secondary constraints in time tell us that no more
constraints appear in the theory. \ Next the constraint classification gives
the following first class
\begin{eqnarray}
\Omega _{\pi }^{(2)} &=&\frac{1}{2}p^{2}\approx 0  \label{p9} \\
\Omega _{k}^{(2)} &=&i\psi _{k}^{\mu }p_{\mu }\approx 0  \label{p10} \\
\Omega _{ik}^{(2)} &=&i\psi _{i}^{\mu }\psi _{k}^{\mu }\approx 0  \label{p11}
\end{eqnarray}
and the second class constraints
\begin{equation}
\Omega _{\mu }^{k}=\pi _{\mu }^{k}-\frac{i}{2}\psi _{\mu }^{k}\approx 0
\label{p12}
\end{equation}
with the help of the second class constraints we construct the Dirac Bracket
(DB) between the canonical variables and obtain
\begin{equation}
\left\{ \psi _{\mu }^{i},\psi _{\nu }^{k}\right\} _{DB}=-i\delta ^{ik}g_{\mu
\nu },\quad \left\{ x_{\mu },p_{\nu }\right\} _{DB}=g_{\mu \nu }  \label{p13}
\end{equation}

\subsection{Invariance}

In the theory with the action (\ref{p4b}), we have three gauge
transformations that do not change their physical sense. The $\tau$%
-reparametrization
\begin{eqnarray}
\delta x &=&\varepsilon \dot{x},\quad \delta \psi =\varepsilon \dot{\psi }
\label{in1} \\
\delta e &=&\left( \varepsilon e\right) ^{.},\quad \delta \chi =\left(
\varepsilon \chi \right) ^{.},\quad \delta f=\left( \varepsilon f\right) ^{.}
\notag
\end{eqnarray}
the invariance under local internal symmetries $O\left( N\right) $%
\begin{eqnarray}
\delta x &=&0,\quad \delta \psi =a\psi  \label{in2a} \\
\delta e &=&0,\quad \delta \chi =a\chi ,\quad \delta f=\dot{a}+af-fa  \notag
\end{eqnarray}
and the invariance under local $\left( n=1\right) $ SUSY transformations
\begin{eqnarray}
\delta x &=&i\alpha \psi ,\quad \delta \psi =e^{-1}\alpha \left( \dot{x}%
-i\chi \psi \right)  \label{in3} \\
\delta e &=&2i\alpha \chi ,\quad \delta \chi =\dot{\alpha }-f\alpha ,\quad
\delta f=0  \notag
\end{eqnarray}

It is interesting to commute two local $\left( n=1\right)$ SUSY
transformations. This gives
\begin{eqnarray}
\left[ \delta _{\alpha },\delta _{\beta }\right] x &=&\delta _{\varepsilon
_{0}}\dot{x}+\delta _{a_{0}}x+\delta _{\alpha _{0}}x  \label{in4} \\
\left[ \delta _{\alpha },\delta _{\beta }\right] \psi &=&\delta
_{\varepsilon _{0}}\dot{\psi }+\delta _{a_{0}}\psi +\delta _{\alpha _{0}}\psi
\\
\left[ \delta _{\alpha },\delta _{\beta }\right] e &=&\delta _{\varepsilon
_{0}}\dot{e}+\delta _{a_{0}}e+\delta _{\alpha _{0}}e \\
\left[ \delta _{\alpha },\delta _{\beta }\right] \chi &=&\delta
_{\varepsilon _{0}}\dot{\chi }+\delta _{a_{0}}\chi +\delta _{\alpha _{0}}\chi
\\
\left[ \delta _{\alpha },\delta _{\beta }\right] f &=&\delta _{\varepsilon
_{0}}\dot{f}+\delta _{a_{0}}f+\delta _{\alpha _{0}}f
\end{eqnarray}
where the new parameters are now field dependent
\begin{equation}
\varepsilon _{0}=2ie^{-1}\alpha \beta ,\quad \alpha _{0}=-\varepsilon
_{0}\chi ,\quad a_{0}=-\varepsilon _{0}f  \label{in5}
\end{equation}
this shows that there is no simple gauge group structure, although the
invariance is still enough to secure good physical properties of the action.

The invariance of the action (\ref{p4b}) is reached if we impose the
conditions at the endpoints for the parameters
\begin{equation}
\varepsilon \left( \tau _{1}\right) =\varepsilon \left( \tau _{2}\right)
=0,\quad \alpha \left( \tau _{1}\right) =\alpha \left( \tau _{2}\right) =0
\label{in6}
\end{equation}

On the other hand it is possible to find the generators of the
transformations (\ref{in1})-(\ref{in3}). We follow the work of Casalbuoni
\cite{Casalbuoni} where the generators of the transformations $F$ are given
by
\begin{equation}
F=p_{a}\delta q^{a}-\varphi ,\quad \delta L=\frac{d\varphi }{d\tau }
\label{in7}
\end{equation}
being $\varphi $ the generating function. To verify the correctness of found
generators we use
\begin{equation}
\delta u=\left\{ u,\epsilon F\right\} _{DB}  \label{in7a}
\end{equation}
where $\epsilon $ is the parameter of a given transformation.

We find for the $\tau$-reparametrizations
\begin{eqnarray}
F &=&i\chi \psi p-\frac{e}{2}p^{2}-\frac{i}{2}f\psi \psi  \label{in8a} \\
\left\{ x^{\mu },\varepsilon F\right\} _{DB} &=&\varepsilon \dot{x}^{\mu
},\quad \left\{ \psi _{k}^{\mu },\varepsilon F\right\} _{DB}=\varepsilon
\dot{\psi }_{k}^{\mu }  \label{in8b}
\end{eqnarray}
internal $O\left( N\right) $ symmetries
\begin{eqnarray}
F_{ik} &=&\frac{i}{2}\psi _{i}^{\mu }\psi _{\mu k}+\chi _{i}\pi _{k}
\label{in9a} \\
\left\{ x^{\mu },aF\right\} _{DB} &=&0,\quad \left\{ \psi _{i}^{\mu
},aF\right\} _{DB}=a_{ik}\psi _{k}^{\mu },\quad \left\{ \chi _{i},aF\right\}
_{DB}=a_{ik}\chi _{k}  \label{in9b}
\end{eqnarray}
and SUSY transformations
\begin{eqnarray}
F_{k} &=&ip_{\mu }\psi _{k}^{\mu }+2i\chi _{k}\pi  \label{in10a} \\
\left\{ x^{\mu },\alpha F\right\} _{DB} &=&i\alpha _{k}\psi _{k}^{\mu
},\quad \left\{ \psi ,\alpha F\right\} _{DB}=e^{-1}\alpha \left( \dot{x}%
-i\chi \psi \right)  \label{in10b} \\
\left\{ e,\alpha F\right\} _{DB} &=&2i\alpha \chi
\end{eqnarray}

To close the invariance we remark that the proposed theory is also invariant
under Poincar\'e transformations, i.e.
\begin{equation}
\delta x^{\mu }=\omega^{\mu }{}_{\nu }x^{\nu }+\epsilon ^{\mu },\quad \delta
\psi _{k}^{\mu }=\omega^{\mu }{}_{\nu }\psi _{k}^{\nu },\quad \delta
e=\delta \chi =\delta f=0  \label{in11}
\end{equation}
with the generators
\begin{equation}
\epsilon ^{a}F_{a}=\epsilon ^{\mu }P_{\mu }+\frac{1}{2}\omega ^{\mu \nu
}M_{\mu \nu }  \label{in12}
\end{equation}
where
\begin{equation}
M_{\mu \nu }=L_{\mu \nu }+S_{\mu \nu },\quad L_{\mu \nu }=x_{\nu }p_{\mu
}-x_{\mu }p_{\nu },\quad S_{\mu \nu }=i\psi _{\mu }^{k}\psi _{\nu }^{k}
\label{in13}
\end{equation}
in this way is constructed the Pauli-Lubanski vector
\begin{equation}
W_{\mu }=\frac{1}{2}\epsilon _{\mu \nu \lambda \rho }P^{\nu }M^{\lambda \rho
},\quad W^{2}=\frac{1}{2}\left( P^{2}S^{2}+2\left( S_{\mu \nu }P^{\nu
}\right) ^{2}\right)  \label{in14}
\end{equation}

\section{Quantization}

The constraint analysis which was done before takes a physical sense when
the quantization is performed and a coherent interpretation of the equation
of motions is given.

With the quantization the canonical variables becomes operators
\begin{equation}
x_{\mu }\rightarrow \widehat{x}_{\mu },\quad p_{\mu }\rightarrow \widehat{p}%
_{\mu },\quad \psi _{\mu }^{i}\rightarrow \widehat{\psi }_{\mu }^{i}
\label{q1}
\end{equation}
and the DB follows the commutator or anticommutator rules
\begin{equation}
\left\{ \widehat{\quad }\right\} \rightarrow i\hbar \left\{ \quad \right\}
_{DB}  \label{q2}
\end{equation}
thus we have the following commutation relations
\begin{equation}
\left\{ \widehat{\psi }_{\mu }^{i},\widehat{\psi }_{\nu }^{k}\right\} =\hbar
\delta ^{ik}g_{\mu \nu },\quad \left[ \widehat{x}_{\mu },\widehat{p}_{\mu }%
\right] =i\hbar g_{\mu \nu }  \label{q3}
\end{equation}

We pick out a general realization for the operator $\widehat{\psi }_{\mu
}^{k}$ satisfying the relation (\ref{q3}) and the equations of motions
\begin{equation}
D\left( \widehat{\psi }_{\mu }^{k}\right) =S\left( Y\right) \left( \left(
\gamma _{5}\right) ^{\otimes \left( k-1\right) }\otimes \gamma _{\mu }\gamma
_{5}\otimes I^{\otimes \left( N-k\right) }\right)  \label{q4}
\end{equation}
here $S(Y)$ is the Young symmetrization operator, $\gamma _{\mu }$ are the
Dirac matrices and $\gamma _{5}$ is given by
\begin{equation}
\gamma _{5}=i\gamma _{0}\gamma _{1}\gamma _{2}\gamma _{3},\quad \left(
\gamma _{5}\right) ^{2}=1  \label{q5}
\end{equation}

The first class constraints are applied into the vector state $\left| \Phi
\right\rangle \equiv \left| \Phi \right\rangle _{\alpha _{1}...\alpha _{N}}$%
. We recall that an internal group symmetry $O(N),$ where $i,k,...=1,2,...,N$%
, is considered in the Lagrangian (\ref{p1}). Thus we obtain
\begin{eqnarray}
p^{2}\left| \Phi \right\rangle _{\alpha _{1}...\alpha _{N}} &=&0  \label{q6}
\\
p^{\mu }\gamma _{\mu }^{k}\left| \Phi \right\rangle _{\alpha _{1}...,\alpha
_{k},...\alpha _{N}} &=&0  \label{q7} \\
\gamma ^{\mu i}\gamma _{\mu }^{k}\left| \Phi \right\rangle _{\alpha
_{1}...,\alpha _{i},...,\alpha _{k},...\alpha _{N}} &=&0  \label{q8}
\end{eqnarray}
the first equation is the mass shell condition in the case of a massless
particle. The second one is a set of linear equations for every Dirac
indices where no symmetrization on the vector state $\left| \Phi
\right\rangle $ is assumed. However when the symmetrization over the vector
state is taken into account, (\ref{q7}) becomes the Bargmann-Wigner \cite%
{Bargmann} equation for a particle with spin $N/2$. The total symmetrical
part of $\left| \Phi \right\rangle $ generates a representation with the
higher spin value. In our case, the third equation is a projector of the
representations of DKP theory, i.e., it separates out a particular spin
representation of the vector state.

In the particular choose: $i,k=1,2,$ i.e. when the internal group symmetry
is $O(2)$, (\ref{q6})-(\ref{q8}) reproduce de DKP equations for massless
particles with spin $0$ and $1$. In this case the realization (\ref{q4})
becomes
\begin{equation}
D\left( \widehat{\psi }_{\mu }^{1}\right) =i\sqrt{\frac{\hbar }{2}}\left(
\gamma _{\mu }\gamma _{5}\otimes 1\right) ,\quad D\left( \widehat{\psi }%
_{\mu }^{2}\right) =i\sqrt{\frac{\hbar }{2}}\left( \gamma _{5}\otimes \gamma
_{\mu }\gamma _{5}\right)  \label{q8a}
\end{equation}

Let's take only two Dirac indices in the vector state $\left\vert \Phi
\right\rangle _{\alpha _{1}\alpha _{2}}$, then using a complete set of Dirac
matrices we decompose $\left\vert \Phi \right\rangle _{\alpha _{1}\alpha
_{2}}$ as follows \cite{Valeri}
\begin{eqnarray}
\left\vert \Phi \right\rangle _{\alpha _{1}\alpha _{2}} &=&a\left( \gamma
^{5}C\right) _{\alpha _{1}\alpha _{2}}\zeta _{5}+a_{1}\left( \gamma
^{5}\gamma ^{\mu }C\right) _{\alpha _{1}\alpha _{2}}\zeta _{5\mu
}+a_{2}C_{\alpha _{1}\alpha _{2}}\zeta  \notag \\
&&+b\left( \gamma ^{\mu }C\right) _{\alpha _{1}\alpha _{2}}\left( \zeta
_{\mu }\right) +b_{1}\left( \Sigma ^{\mu \nu }C\right) _{\alpha _{1}\alpha
_{2}}\zeta _{\mu \nu }  \label{q9}
\end{eqnarray}
here $a,a_{1},b,b_{1}$ and $a_{2}$ must be considered as free parameters and
will be adjusted to assure the correctness of the final equations. The term:
$C_{\alpha _{1}\alpha _{2}}\zeta $, is referred to a trivial representation
and we do not consider it, therefore, we set $a_{2}=0.$

We also have
\begin{equation}
\Sigma ^{\mu \nu }=\frac{1}{2}\left( \gamma ^{\mu }\gamma ^{\nu }-\gamma
^{\nu }\gamma ^{\mu }\right)  \label{q10}
\end{equation}
and $C$ is the charge conjugation matrix
\begin{equation}
C^{T}=-C.  \label{q11}
\end{equation}
Considering the properties of the matrix $C$ we obtain the antisymmetrical
\begin{equation}
\left\vert \Phi \right\rangle _{\left[ \alpha _{1}\alpha _{2}\right]
}=a~\left( \gamma ^{5}C\right) _{\alpha _{1}\alpha _{2}}~\zeta
_{5}+a_{1}\left( \gamma ^{5}\gamma ^{\mu }C\right) _{\alpha _{1}\alpha
_{2}}\zeta _{5\mu }  \label{q12}
\end{equation}
and the symmetrical part of the vector state.
\begin{equation}
\left\vert \Phi \right\rangle _{\left\{ \alpha _{1}\alpha _{2}\right\}
}=b\left( \gamma ^{\mu }C\right) _{\alpha _{1}\alpha _{2}}\zeta _{\mu
}+b_{1}\left( \Sigma ^{\mu \nu }C\right) _{\alpha _{1}\alpha _{2}}\zeta
_{\mu \nu }.  \label{q13}
\end{equation}

\bigskip

Thus for the particular case of $O(2)$ symmetry we obtain
\begin{eqnarray}
p^{2}\left\vert \Phi \right\rangle _{\alpha _{1}\alpha _{2}} &=&0
\label{q14} \\
p^{\mu }\gamma _{\mu }^{(1)}\left\vert \Phi \right\rangle _{\alpha
_{1}\alpha _{2}} &=&0,\quad p^{\mu }\gamma _{\mu }^{(2)}\left\vert \Phi
\right\rangle _{\alpha _{1}\alpha _{2}}=0  \label{q15} \\
\gamma _{\mu }^{(1)}\gamma ^{\mu (2)}\left\vert \Phi \right\rangle _{\alpha
_{1}\alpha _{2}} &=&0  \label{q16}
\end{eqnarray}
these relations give the DKP equation for spin $0$ and spin $1$. The
relation (\ref{q16}) can be shown to be a projector that separates the
corresponding sector of the vector state $\left| \Phi \right\rangle _{\alpha
_{1}\alpha _{2}}.$

\subsection{Spin 0}

Let's take the antisymmetrical part of the vector state $\left\vert \Phi
\right\rangle _{\alpha _{1}\alpha _{2}}$ and replace it in one of the
equations (\ref{q15}), then we obtain
\begin{equation}
\left( p_{\mu }\gamma ^{\mu }\right) _{\alpha \alpha _{1}}\left\vert \Phi
\right\rangle _{\left[ \alpha _{1}\alpha _{2}\right] }=\left( p_{\mu }\gamma
^{\mu }\right) _{\alpha \alpha _{1}}\left[ a\left( \gamma ^{5}C\right)
_{\alpha _{1}\alpha _{2}}\zeta _{5}+a_{1}\left( \gamma ^{5}\gamma ^{\nu
}C\right) _{\alpha _{1}\alpha _{2}}\zeta _{5\nu }\right] =0  \label{s1}
\end{equation}
multiplying on the right side by $\left( C^{-1}\gamma ^{5}\right) _{\alpha
_{2}\alpha }$ and considering $\gamma _{5}^{2}=1$, we have
\begin{equation}
p_{\mu }\left[ a\left( \gamma ^{\mu }\right) _{\alpha \alpha }\zeta
_{5}-a_{1}\left( \gamma ^{\mu }\gamma ^{\nu }\right) _{\alpha \alpha }\zeta
_{5\nu }\right] =0  \label{s2}
\end{equation}
with the use of the trace properties the equation (\ref{s2}) results in
\begin{equation}
a_{1}\left( p^{\mu }\zeta _{5\mu }\right) =0  \label{s3}
\end{equation}

On the other hand, if we multiply the equation (\ref{s1}) by $\left(
C^{-1}\gamma ^{5}\gamma ^{\lambda }\gamma ^{\rho }\right) _{\alpha
_{2}\alpha }$ and taking the trace operation we got to
\begin{equation}
a_{1}\left( p^{\mu }\zeta _{5}^{\nu }-p^{\nu }\zeta _{5}^{\mu }\right) =0
\label{s4}
\end{equation}
for $a_{1}\neq 0$, one solution for the last relation is given by
\begin{equation}
\zeta _{5}^{\mu }=p^{\mu }\zeta _{5}  \label{s5}
\end{equation}
Thus equations (\ref{s3}) and (\ref{s4}) are the equations for the spin $0$
particles and the equation (\ref{s3}) gives the massless Klein-Gordon
equation for the scalar field $\zeta _{5}$.

Now if we multiply (\ref{s1}) on the right side by $\left( C^{-1}\gamma
^{5}\gamma ^{\lambda }\right) _{\alpha _{2}\alpha }$ we obtain
\begin{equation}
p_{\mu }\left[ a\left( \gamma ^{\lambda }\gamma ^{\mu }\right) _{\alpha
\alpha }\zeta _{5}-a_{1}\left( \gamma ^{\lambda }\gamma ^{\mu }\gamma ^{\nu
}\right) _{\alpha \alpha }\zeta _{5\nu }\right] =0  \label{s6}
\end{equation}
using again the trace properties for the Dirac matrices a third relation is
obtained
\begin{equation}
a\left( p^{\mu }\zeta _{5}\right) =0  \label{s7}
\end{equation}
this equation is compatible with the equation (\ref{s3}) and (\ref{s4}) if
only if $a=0.~$

\subsection{Spin 1}

Now we take the symmetrical part of the vector state $\left\vert \Phi
\right\rangle _{\alpha _{1}\alpha _{2}}$, the equation (\ref{q15}) becomes
\begin{equation}
\left( p_{\mu }\gamma ^{\mu }\right) _{\alpha \alpha _{1}}\left[ b\left(
\gamma ^{\nu }C\right) _{\alpha _{1}\alpha _{2}}\zeta _{\nu }+b_{1}\left(
\Sigma ^{\nu \lambda }C\right) _{\alpha _{1}\alpha _{2}}\zeta _{\nu \lambda }%
\right] =0.  \label{sp1}
\end{equation}

\bigskip Multiplying on the right side by~ $\left( C^{-1}\gamma ^{\rho
}\right) _{\alpha _{2}\alpha }$ we get
\begin{equation}
p_{\mu }\left[ \left( \gamma ^{\mu }\gamma ^{\nu }\gamma ^{\rho }\right)
_{\alpha \alpha }\zeta _{\nu }+\left( \gamma ^{\mu }\Sigma ^{\nu \lambda
}\gamma ^{\rho }\right) _{\alpha \alpha }\zeta _{\nu \lambda }\right] =0
\label{sp4}
\end{equation}
using the trace properties for the $\gamma ^{\mu }$-matrices it simplifies
to give
\begin{equation}
b_{1}\left( p^{\lambda }\zeta _{\lambda \rho }\right) =0  \label{sp5}
\end{equation}

Multiplying (\ref{sp1}) by $\left( C^{-1}\gamma ^{\rho }\gamma ^{\sigma
}\gamma ^{\tau }\right) _{\alpha _{2}\alpha }$ it simplifies to be
\begin{equation}
p_{\mu }\left[ \left( \gamma ^{\mu }\gamma ^{\nu }\gamma ^{\rho }\gamma
^{\sigma }\gamma ^{\tau }\right) _{\alpha \alpha }\zeta _{\nu }+\left(
\gamma ^{\mu }\Sigma ^{\nu \lambda }\gamma ^{\rho }\gamma ^{\sigma }\gamma
^{\tau }\right) _{\alpha \alpha }\zeta _{\nu \lambda }\right] =0  \label{m55}
\end{equation}
tracing the equation above and considering the antisymmetric character of
the tensor field\ $\zeta ^{\rho \tau }$ we get the Bianchi relation
\begin{equation}
b_{1}\left( p^{\rho }\zeta ^{\tau \sigma }+p^{\sigma }\zeta ^{\rho \tau
}+p^{\tau }\zeta ^{\sigma \rho }\right) =0  \label{m56}
\end{equation}
If we set $b_{1}\neq 0$, one possible solution of the relation
(\ref{m56}) can be obtained if we put
\begin{equation}
\zeta ^{\mu \nu }=p^{\mu }\zeta ^{\nu }-p^{\nu }\zeta ^{\mu }  \label{m57}
\end{equation}
i.e. the strength tensor of the Maxwell theory and the equation (\ref{sp5})
becomes the Maxwell equation for the electromagnetic field .

\bigskip We can obtain more two equations:\ the first one is gotten
multiplying (\ref{sp1}) on the right side by $\left( C^{-1}\right) _{\alpha
_{2}\alpha }$ we have
\begin{equation}
bp_{\mu }\left( \gamma ^{\mu }\gamma ^{\nu }\right) _{\alpha \alpha }\zeta
_{\nu }=0  \label{sp2}
\end{equation}
with the help of the trace properties for the Dirac matrices we obtain
\begin{equation}
b\left( p_{\mu }\zeta ^{\mu }\right) =0,  \label{sp3}
\end{equation}
to get the second one we multiply (\ref{sp1}) by $\left( C^{-1}\gamma ^{\rho
}\gamma ^{\sigma }\right) _{\alpha _{2}\alpha }$ and next we take the trace
operation over the $\gamma ^{\mu }$-matrices to obtain
\begin{equation}
b\left( p^{\mu }\zeta ^{\nu }-p^{\nu }\zeta ^{\mu }\right) =0  \label{sp6}
\end{equation}
The equations (\ref{sp3}) and (\ref{sp6}) are compatible with the equations (%
\ref{sp5}), (\ref{m56}) and (\ref{m57}) if and only if we set $b=0$.

\subsection{Topological solutions}

On the other hand we can get two additional solutions if we set $b\neq 0$
and $b_{1}=0$. Thus the first solution is getting when we solve the equation
(\ref{sp3}) choosing
\begin{equation}
\zeta ^{\mu }=p_{\nu }\zeta ^{\mu \nu }  \label{ts-1}
\end{equation}
where $\zeta ^{\mu \nu }$ is an antisymmetrical tensor field satisfying the
equation (\ref{sp6}). \

\bigskip

And the second solution is founded when set the vector field in the equation
(\ref{sp3}) being
\begin{equation}
\zeta ^{\mu }=\epsilon ^{\mu \nu \alpha \beta }p_{\nu }\zeta _{\alpha \beta }
\label{ts-2}
\end{equation}

The equations (\ref{ts-1}) and (\ref{ts-2}) are topological field solutions
for the spin 1 and spin 0 sectors \cite{buchbinder}, respectively. Such
topological solutions were found in the massless DKP theory by
Harish-Chandra \cite{reft0} and in the context of usual Klein-Gordon and
Maxwell theories studying their higher tensor representations by Deser and
Witten \cite{reft1} and Townsend \cite{reft2}.

\section{Superspace Formulation}

As a natural way we extend the previous analysis of the action and give the
formulation in terms of superspace.

Firstly we consider the motion of the particle in the large superspace (big
SUSY) $\left( X_{\mu },\Theta _{\alpha }\right) $\footnote{%
When the interaction is switched on, we must to include a complex grassmann
spinor field $\overline{\Theta }_{\overset{.}{\alpha }}.$ This enable us to
consider theories with interacting charged particles.} whose trajectory is
parametrized by the proper supertime $\left( \tau ,\eta _{1},\eta
_{2}\right) $ of dimension $\left( 1/2\right) $, here $\eta _{1},\eta _{2}$\
are the grassmann real superpartners of the convencional time $\tau $. In
this way the coordinates of the particle are scalar superfields in the
little superspace (little SUSY). For this case we have\footnote{%
We recall that this form is valid only for the case of two indices $i=1,2$.
If we want to analyse theories with a bigger internal symmetry $O\left(
N\right) ,$ we need to include a more terms.}
\begin{eqnarray}
X_{\mu }\left( \tau ,\eta _{1},\eta _{2}\right) &=&x_{\mu }\left( \tau
\right) +i\eta _{i}\psi _{\mu }^{i}\left( \tau \right) +i\eta _{i}\eta
_{j}F_{\mu }^{ij}\left( \tau \right)  \label{su1} \\
\Theta _{\alpha }\left( \tau ,\eta _{1},\eta _{2}\right) &=&\theta _{\alpha
}\left( \tau \right) +\eta _{i}\lambda _{\alpha }^{i}\left( \tau \right)
+\eta _{i}\eta _{j}\mathcal{F}_{\alpha }^{ij}\left( \tau \right)  \label{su2}
\end{eqnarray}
where $i,j=1,2;$ $\psi _{\mu }^{i}$ is the grassman superpartner of the
common coordinate $x_{\mu };$ $\lambda _{\alpha }^{i}$\ is a commuting
majorana spinor, superpartner of the grassmann variables $\theta _{\alpha }$%
. $F_{\mu }^{ij}=-F_{\mu }^{ji}$ and $\mathcal{F}_{\alpha }^{ij}=-\mathcal{F}%
_{\alpha }^{ji}$ are antisymmetric fields.

In order to construct an action which is invariant under general
transformations in superspace we introduce the supereinbein $E_{M}^{A}\left(
\tau ,\eta _{1},\eta _{2}\right) $, where $M$ [$A$] are a curved [tangent]
indices and $D_{A}=E_{A}^{M}\partial _{M}$ is the supercovariant general
derivatives, here $E_{A}^{M}$ is the inverse of $E_{M}^{A}$. If we take a
special gauge
\begin{equation}
E_{M}^{\alpha }=\Lambda \overline{E}_{M}^{\alpha },\quad E_{M}^{a}=\Lambda
^{1/2}\overline{E}_{M}^{a}  \label{su3}
\end{equation}
where
\begin{equation}
\overline{E}_{\mu }^{\alpha }=1,\quad \overline{E}_{\mu }^{a}=0,\quad
\overline{E}_{m}^{\alpha }=-i\eta ,\quad \overline{E}_{m}^{a}=1  \label{su4}
\end{equation}
is the flat space supereinbein, then the superscalar field $\Lambda $ an the
derivative $D_{A}$ takes the form
\begin{eqnarray}
\Lambda \left( \tau ,\eta _{1},\eta _{2}\right) &=&e\left( \tau \right)
+i\eta _{i}\chi _{i}\left( \tau \right) +i\eta _{i}\eta _{j}f_{ij}\left(
\tau \right) ,  \label{su5} \\
\quad \overline{D}_{a} &\equiv &D_{i}=\frac{\partial }{\partial \eta ^{i}}%
+i\eta _{i}\frac{\partial }{\partial \tau },\quad \overline{D}_{\alpha
}=\partial _{\tau }  \label{su6}
\end{eqnarray}
here $e\left( \tau \right) $ is the graviton field and $\chi _{i}\left( \tau
\right) $ the gravitino field of the two-dimensional $n=2$ supergravity; $%
f_{ij}=-f_{ji}$ is an antisymmetric matrix field. It is no difficult to
prove that $\left( \overline{D}_{a}\right) ^{2}\equiv \left( D_{i}\right)
^{2}=i\partial _{\tau }$

In this way the extension to superspace of the action (\ref{p4b}), is given
by\footnote{%
The presence of the superscalar field $\Lambda $ is to guarantee the local
SUSY invariance.}
\begin{equation}
S=\frac{1}{4}\int d\tau d\eta _{1}d\eta _{2}\Lambda ^{-1}\epsilon
_{ij}D_{i}X_{\mu }D_{j}X^{\mu }  \label{su7}
\end{equation}%
here \ $\epsilon _{ij}$ is the antisymmetric matrix: $\epsilon
_{12}=-\epsilon _{21}=1,$ $\epsilon _{11}=\epsilon _{22}=0$. Using the
property $\Lambda \Lambda ^{-1}=1$ for the supereinbein field we obtain
\begin{eqnarray}
\Lambda ^{-1}\left( \tau ,\eta _{1},\eta _{2}\right) &=&e^{-1}\left( \tau
\right) -ie^{-2}\left( \tau \right) \eta _{i}\chi _{i}\left( \tau \right)
-ie^{-2}\left( \tau \right) \eta _{i}\eta _{j}f_{ij}\left( \tau \right)
\notag \\
&&+e^{-3}\left( \tau \right) \eta _{i}\eta _{j}\chi _{i}\left( \tau \right)
\chi _{j}\left( \tau \right)  \label{su8}
\end{eqnarray}%
After some manipulations and integrating over the grassmann variables we
have
\begin{eqnarray}
S &=&\int d\tau \left( -\frac{1}{2}e^{-1}\overset{.}{x}^{2}+\frac{i}{2}%
e^{-1}\psi _{i}\overset{.}{\psi }_{i}+\frac{i}{2}e^{-2}\chi _{i}\psi _{i}%
\overset{.}{x}+\frac{i}{2}e^{-2}f_{ij}\psi _{i}\psi _{j}\right.  \notag \\
&&\left. +\frac{1}{2}e^{-3}\chi _{i}\psi _{i}\chi _{j}\psi
_{j}+e^{-1}F^{2}-ie^{-2}F_{ij}\chi _{i}\psi _{j}\right)  \label{su9}
\end{eqnarray}%
redefining the fields
\begin{equation}
\chi =e^{1/2}\chi ^{\prime },\quad \psi =e^{1/2}\psi ^{\prime },\quad
f=ef^{\prime },\quad F=eF^{\prime }  \label{su10}
\end{equation}%
we obtain
\begin{eqnarray}
S &=&\int d\tau \left( -\frac{1}{2}e^{-1}\overset{.}{x}^{2}+\frac{i}{2}\psi
_{i}\overset{.}{\psi }_{i}+\frac{i}{2}e^{-1}\chi _{i}\psi _{i}\overset{.}{x}+%
\frac{i}{2}f_{ij}\psi _{i}\psi _{j}\right.  \notag \\
&&\left. +\frac{1}{2}e^{-1}\chi _{i}\psi _{i}\chi _{j}\psi
_{j}+eF^{2}-iF_{ij}\chi _{i}\psi _{j}\right)  \label{su11}
\end{eqnarray}%
we see that this action is identical to the proposed in (\ref{p4b}) when we
put $F=\chi \psi $, i.e. when the fermion coordinate and the gravitino field
are coupled.

This shows that considering the correct inclusion of internal symmetries in
the superspace formulation we obtain, in the special case, the same action
proposed from the pseudoclassical point of view. The internal symmetry group
$O\left( N\right) $ is connected to the number of grassmann variables $\eta
_{i}.$

\section{Conclusions}

In this work we give an action for the massless DKP theory by using
Grassmann variables and the consistence of the equations of motions are
assured by means of the inclusion of boundary terms. We also verified the
invariance under $\tau $-reparametrizations, local SUSY and internal group $%
O(N)$ transformations, the generators of these transformations are
also found. We carried out the constraint analysis of the theory and
verified that after quantization a possible inconsistency can
appear, nevertheless the further analysis allow us to solve it with
the introduction of some parameters that play a role of regulators
of the theory. By the way an important result in this context was
obtained, i.e. an additional topological solution for the spin 0 and
1 is derived from this model. As a natural continuation of the
presented action we extended the studies to superspace formalism
obtaining under some conditions the same initial pseudoclassical
action.

For the further development of the theory we are working to accomplish the
analysis through the most powerful method for a theory with constraints,
i.e. via the BFV-BRST method, which can open the possibility of calculating
the propagator of the resulting theory using the path integral
representation. And, for further studies the inclusion of interactions
(i.e., electromagnetic, Yang-Mills and gravitational fields) in the theory
will be discussed.

\subsection*{Acknowledgements}

RC (grant 01/12611-7) and MP thank FAPESP and CAPES for full
support, respectively, BMP thanks CNPq and FAPESP (grant 02/00222-9)
for partial support, JSV thanks FAPESP (grant 00/03812-6) and
FAPEMIG (grant 00193/06) for partial and full support, respectively.

\end{document}